\documentclass[epjc3,pdftex,twocolumn,floatfix]{svjour3}      
\usepackage{inputenc,amsmath,amssymb,bm,graphicx,tikz,cite} 
\usepackage{xcolor}
\usepackage{orcidlink} 

\usepackage{graphicx}
\usepackage{dcolumn}
\usepackage{bm}
\usepackage{tikz}
\usepackage{ulem}

\usepackage{algorithm2e}
\RequirePackage[T1]{fontenc}

\smartqed  

\RequirePackage{graphicx}
\RequirePackage{mathptmx}      
\RequirePackage{flushend}
\RequirePackage[numbers,sort&compress]{natbib}

\journalname{Eur. Phys. J. C}

\usetikzlibrary{decorations.pathmorphing,patterns}

\begin{document}


\title{Isospin kaon anomaly and its consequences }

\author{Francesco Giacosa\thanksref{addr1}\orcidlink{0000-0002-7290-9366} \and Martin Rohrmoser\thanksref{addr1}\orcidlink{0000-0003-2311-832X} }
 \institute{Institute of Physics, Jan Kochanowski University, ul. Uniwersytecka 7, Kielce, Poland\label{addr1} }
\maketitle
\begin{abstract}
Isospin symmetry is well fulfilled in the QCD vacuum, as
evidenced by small mass differences of isospin partners and suppressed isospin-violating decays. 
Recently, the NA61/SHINE
collaboration reported an unexpectedly large isospin-violating
charged-to-neutral kaon ratio in Ar-Sc heavy-ion collisions (HIC).
Using a quark recombination approach, 
we introduce a function of kaon multiplicities that reduces to unity in the isospin-symmetric limit independently of the scattering energy and type of nuclei. Using this quantity,
we show that nucleus-nucleus collisions violate isospin sizably
(at the $6.4\sigma$--level), while proton-proton data on kaon multiplicities do not.
We  predict other
isospin-violating enhancements in HIC, such as the proton-to-neutron ratio $p/n \sim 1.2$ and the hyperon ratio $\Sigma^{+}%
/\Sigma^{-}\sim1.4$. Finally, we extend the approach to antiquarks in the
initial state, useful for e.g. pion-nucleus scattering reactions. 
\end{abstract}
%
%
The isospin symmetry, introduced by Heisenberg in 1932 to combine the almost
mass degenerate proton and neutron into a unique object, the nucleon,
is the first example of an internal symmetry in particle physics \cite{Heisenberg:1932dw}. The name isospin, given by Wigner shortly after \cite{Wigner:1936dx}, arises from the mathematical
similarity to the spin: both are based on the special unitary group $SU(2)$. The proton $p$ and
the neutron $n$ form an isospin multiplet, specifically an
isodoublet with total isospin $I=1/2$ and component $I_{z}=1/2$ for
$p$ and $I_{z}=-1/2$ for $n$. The nuclear force is invariant under isospin transformations (i.e. $SU(2)$-rotation in the $(p,n)$-space\footnote{More specifically,  an isospin transformation amounts to  $(p,n)^{T}\rightarrow U(p,n)^{T}$ where $U$ is an $SU(2)$ matrix. The charge-symmetry transformation, given by $U_{ij} = \epsilon_{ij}$, switches $p$ with $n$. Similar relations hold for other isospin multiplets.}),
enabling one to classify and understand nuclei \cite{2006NatPh...2..311W,Webb}.

Just as the proton ($uud$) and the neutron ($udd$), the two kaon
states $K^{+}=u\bar{s}$ and $K^{0}=d\bar{s}$ form an isodoublet
($I=1/2$ and $I_z= \pm 1/2$). 
The two kaons above form the most evident
realization of an isospin multiplet because the $s$-quark is `isospin-inert' (invariant): in simple
terms $K^{+}$and $K^{0}$ correspond to $(u,d)^{T}$.
The other two kaonic
states $\bar{K}^{0} =- s\bar{d}$ and $K^{-} = s \bar{u}$ form an isodoublet which corresponds to the antiquark states
$(-\bar{d},\bar{u})$.  For these kaonic states,  isospin transformations are equivalent to rotations in the $(K^+,K^0)$ and $(-\bar{K}^0,K^-)$ spaces, respectively. 

The charge-symmetry transformation, denoted as $C_I$, is an important specific isospin transformation that, in general, swaps two isomultiplet members with $\pm I_z$, such as: $p \leftrightarrow n$, $K^+ \leftrightarrow K^0$, $K^0 \leftrightarrow K^-$. The lightest hadron, the pion, forms an isotriplet $\pi^+$, $\pi^0$, and $\pi^-$. Under a $C_I$-transformation, \\$\pi^+ \leftrightarrow \pi^-$.

Isospin symmetry is not exact because the quarks $u$ and $d$ are not exactly interchangeable, i.e., they are not
degenerate in mass ($m_u \neq m_d$). Nevertheless, isospin symmetry is very well fulfilled in the QCD phenomenology. 
Isospin breaking is visible in small mass differences, \\e.g.  $\left(  m_{K^{+}}-m_{K^{0}}\right)  /\left(m_{K^{+}}+m_{K^{0}}\right)  \simeq -0.004$.
Isospin-violating decays exist but are typically suppressed. An example is the decay $\eta^{\prime} \rightarrow \pi^+ \pi^- \pi^{0}$ with a small branching ratio of $3.61 \cdot 10^{-4}$ \cite{ParticleDataGroup:2024cfk}. This branching ratio is proportional to $(m_d-m_u)^2$ \cite{Serpukhov-Brussels-AnnecyLAPP:1984xib}.
Another example is the isospin-breaking $\omega$-$\rho$ mixing \cite{OConnell:1995nse}. The small related decays $\omega \rightarrow \pi^+ \pi^-$ and $\rho \rightarrow \pi^+ \pi^- \pi^{0}$ break the so-called $G$-parity. $G$-parity is charge conjugation (particle-antiparticle switch $\mathcal{C}$) \textit{and} charge-symmetry ($I_z$-switch $C_I$), in formulas: $G = \mathcal{C} \cdot C_I $. Since charge conjugation $\mathcal{C}$ is exactly fulfilled in strong and electromagnetic interactions, breaking $G$-parity implies breaking of charge symmetry and, hence, of isospin. In the PDG \cite{ParticleDataGroup:2024cfk}, $G$-parity is reported for all mesons with integer isospin.
See, for instance, the results of the extended Linear Sigma model \cite{Parganlija:2012fy} for a variety of isospin-violating decays in Refs. \cite{Kovacs:2024cdb,Giacosa:2024epf}.

Finally, pion-pion, pion-kaon, and pion-nucleon scattering fulfill isospin symmetry \cite{Pennington:2005be}, but a suppressed breaking has been spotted, e.g. Ref. \cite{Hoferichter:2009ez}.

The results quoted above show that isospin breaking is always small ($\lesssim  5\%$).
One then expects that isospin symmetry is fulfilled at the same level of accuracy in heavy-ion collisions. In particular, if the initial states of two colliding nuclei contain an equal amount of neutrons and protons, $Q/A = 1/2$, the corresponding ensemble of events is charge-symmetry invariant. As a consequence, the average
 multiplicities for $C_I$-partners are equal, e.g.: $\langle K^{+} \rangle = \langle K^{0} \rangle $, $\langle \bar{K}^{0} \rangle =\langle  K^{-} \rangle$, $\langle p \rangle = \langle n \rangle$, $\langle \pi^+ \rangle = \langle \pi^- \rangle $, etc.
\cite{Brylinski:2023nrb,Giacosa:2024bup}.
Thus, the charged-vs-neutral ratio of kaon multiplicites
\begin{equation}
R_{K}=\frac{\left\langle K^{+}\right\rangle +\left\langle K^{-}\right\rangle
}{\left\langle K^{0}\right\rangle +\left\langle \bar{K}^{0}\right\rangle
}=\frac{\left\langle K^{+}\right\rangle +\left\langle K^{-}\right\rangle
}{\left\langle 2K_{S}^{0}\right\rangle }%
\label{rk1}
\end{equation}
reduces to unity, $R_K =1$, for $Q/A =1/2$ in the isospin-symmetric limit
\footnote{The $K_S^0$ state reads $\left\vert K_{S}^{0}\right\rangle =(\left\vert K^{0}\right\rangle -\left\vert
\bar{K}^{0}\right\rangle )/\sqrt{2}$, leading to 
$\left\langle K^{0}%
\right\rangle +\left\langle \bar{K}^{0}\right\rangle =\left\langle 2K_{S}%
^{0}\right\rangle $.
} 
.

Surprisingly, the recent measurement of Ar-Sc scattering by the NA61/SHINE experiment at CERN \cite{NA61SHINE:2023azp} reports $R_K = 1.184 \pm 0.061$. 
The initial system corresponds to
$Q/A = 0.458$, quite close to  $1/2$. The well-known hadron resonance gas (HRG) model \cite{Vovchenko:2018fmh,Vovchenko:2019pjl,Torrieri:2004zz}, which contains known isospin-breaking effects (such as resonance decays, e.g. the $\phi$-meson) leads to $R_K \simeq 1.04$, and thus cannot reproduce the NA61/SHINE experimental outcome. This is also true for the compilation of previous results from heavy-ion experiments, resulting in an overall theory-experiment mismatch of about $4.7 \sigma$ \cite{NA61SHINE:2023azp}. The HRG results have been cross-checked by the UrQMD approach \cite{Bleicher:1999xi,Bleicher:2022kcu}.  Recently, the UrQMD approach has been updated by modifying the string fragmentation into quark-antiquark pairs  ~\cite{Reichert:2025znn}. Requiring that $\bar{u}u$ is produced three times more often than the $\bar{d}d$ one, both $e^+e^-$ and $AA$ data on $R_K$ can be described. In this respect, this result confirms that a sizable isospin-symmetry breaking is required.

In this work, we intend to investigate further the `kaon anomaly' and present novel predictions that result from it. To this end, we consider a simple, effective treatment for comparing multiplicities; see Refs. \cite{Stepaniak:2023pvo,Bonesini:2001iz}. 
This is a quark recombination model\footnote{As reported in Ref. \cite{Bonesini:2001iz} the model was developed by N. Doble, L. Gatignon, P. Grafstrom, NA31 Internal note
83 (1990). According to the authors, the formula and its derivation are due to Horst Wachsmuth.}, which
we briefly recapitulate and extend below. We prefer to present the model in terms of constituent quarks, rather than using the valence/sea quark terminology, which is more appropriate for describing the internal structure of hadrons, such as the proton\footnote{As is well known, the proton exhibits a rich structure in terms of sea and valence quarks; see, e.g., Refs.~\cite{SeaQuest:2021zxb,Geesaman:2018ixo,Chang:2014jba}. }. For our counting scheme, the proton is made of two constituent quarks $u$ and one constituent $d$. Clearly, the constituent quark is far from being a simple object (it may be seen as a valence quark surrounded by gluons and valence sea quarks, e.g. \cite{Gluck:1997ww,Boffi:2002yy}), but is sufficient for our counting. 
The initial state of two colliding nuclei contains a certain number of inital or `preexisting' constituent 
quarks $u$ and $d$, which we denote as $n_{u}$ and $n_{d}$. For instance, proton-proton scattering amounts to $n_u = 4$ and $n_d = 2$. 
As a result of the collisions, quark-antiquark constituent pairs are created by the QCD vacuum. We denote them as
$\alpha=n_{u}^{vacuum}=n_{\bar{u}}^{vacuum},$ $\beta=n_{d}^{vacuum}=n_{\bar{d}}^{vacuum},$
and $\gamma=n_{s}^{vacuum}=n_{\bar{s}}^{vacuum}.$ 
In this respect, the creation of constituent quark-antiquark pairs is reminiscent of the $^3P_0$ model used to describe mesonic decays in constituent quark models, see e.g. \cite{Amsler:1995td,Ackleh:1996yt} and refs. therein. 
Isospin symmetry implies
$\alpha=\beta,$ while flavor symmetry means $\alpha=\beta=\gamma$. 
The total
number of (anti)quarks is $n_{tot}=n_{u}+n_{d}+2\alpha+2\beta+2\gamma.$ 
Hence, the probability that a quark picked randomly out of this ensemble of collisions is of the type $u$ amounts to
\begin{equation}
p(u)=\frac{n_{u}+\alpha}{n_{tot}}=p_{initial}(u)+p_{vacuum}(u)
\text{ ,}
\end{equation}
with $p_{initial}(u)=n_{u}/n_{tot}$ and $p_{vacuum}(u)=\alpha/n_{tot}.$ Similar
relations hold for the other (anti)quarks.

Next, we consider a pair of (anti)quarks. Within the quark recombination scheme, quarks are treated as uncorrelated: the probability that a quark pair converts into a
meson $K^{+}$ is proportional to the probability that one quark is of the type
$u$ multiplied by the probability that the second quark is of the type
$\bar{s}$, leading to the probability $p(u)p(\bar{s})$. We thus interpret the $K^+$ as built of one constituent quark and one constituent antiquark. 
For the four kaon types, we obtain:
\begin{align}
p(K^{+})  &  \propto n_{u}\gamma+\alpha\gamma\text{ ; }p(K^{-})\propto
\alpha\gamma\text{ } \label{kpm} \text{ ;}\\
p(K^{0})  &  \propto n_{d}\gamma+\beta\gamma\text{; }p(\bar{K}^{0}%
)\propto\beta\gamma \text{ .}\label{k0b}
\end{align}
Within this approach, the ratio $R_{K}$ of Eq. (\ref{rk1}) reads
\begin{equation}
R_{K}= \frac{\left\langle K^{+}\right\rangle +\left\langle K^{-}\right\rangle
}{\left\langle 2K_{S}^{0}\right\rangle} 
 = \frac{n_{u}+2\alpha}{n_{d}+2\beta}
\text{ .}
\end{equation}
In the isospin-symmetric limit ($\alpha=\beta$), $R_{K}=1$ if $n_{u}=n_{d}$, which corresponds to 
$Q/A=1/2$. 
We thus recover the results $R_K = 1$ obtained using charge-symmetry arguments \cite{NA61SHINE:2023azp}. 
For other choices of $n_u$ and $n_d$ (and thus of $Q/A$), the ratio $R_{K}$ is not as simple. The vacuum/initial quark ratio $\alpha/n_{u}$ is in general, energy dependent; see below.

Two additional general consequences for the multiplicities arise from Eqs. (\ref{kpm}) and (\ref{k0b}):
\begin{equation}
\left\langle K^{-}\right\rangle =\left\langle \bar{K}^{0}\right\rangle \text{
; }\frac{\left\langle K^{0}\right\rangle -\left\langle \bar{K}^{0}%
\right\rangle }{\left\langle K^{+}\right\rangle -\left\langle K^{-}%
\right\rangle }=\frac{n_{d}}{n_{u}}=\frac{2-\frac{Q}{A}}{1+\frac{Q}{A}}\text{
.}
\label{constraints}
\end{equation}
These equations cannot
be directly checked experimentally because  $\left\langle \bar{K}^{0}\right\rangle $ and
$\left\langle K^{0}\right\rangle $ are not measured independently.
They can be however used to show that the following ratio
of multiplicities:
\begin{equation}
\tilde{R}_{K}=R_{K}+\left(  \frac{1-2\frac{Q}{A}}{1+\frac{Q}{A}}\right)
\frac{\left\langle K^{+}\right\rangle -\left\langle K^{-}\right\rangle
}{2\left\langle K_{S}^{0}\right\rangle }=\frac{n_{d}+2\alpha}{n_{d}+2\beta}
\text{ ,}
\label{rkt}
\end{equation}
gives unity, $\tilde{R}_K = 1$, in the isospin-conserved limit ($\alpha = \beta$),
independently of the energy of the collisions, as in $R_{K}=1$
for the $Q/A=1/2$ case. In general, $\tilde{R}_K$ depends also on the difference $\langle K^+ \rangle - \langle K^- \rangle$, but for $Q/A=1/2$ it reduces to $\tilde{R}%
_{K}=R_{K}$.

\begin{figure}
    \centering
    \includegraphics[width=\linewidth,clip=true,trim=7pt 4pt 0 0]{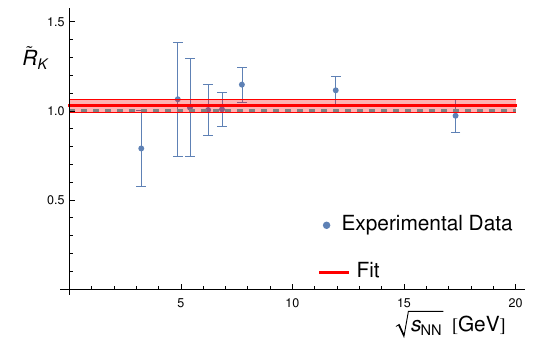}
    \caption{Experimental data for $\tilde{R}_K$ of Eq. (\ref{rkt}) from proton-proton collisions, together with the weighted average (constant fit) $\tilde{R}_K =1.030\pm0.038$. The dashed line is the isospin-symmetric prediction $\tilde{R}_K = 1$.}
    \label{fig:rktildepp}
\end{figure}

For \textit{pp} collisions $Q/A=1$, hence the constrain $\tilde{R}_{K}=1$ delivers:
\begin{equation}
\left\langle K^{+}\right\rangle +3\left\langle K^{-}\right\rangle
=4\left\langle K_{S}^{0}\right\rangle
\text{ .}
\end{equation}
This relation has been discussed in Ref. \cite{Stepaniak:2023pvo}, where it is shown that it qualitatively describes  \textit{pp} data, while the relation $K^++K^- = K^0 +\bar{K}^0$ does not. 

To be rigorous about this important point,  
we consider the $\chi^2$-function 
\begin{equation}
\chi^{2}(y)=\sum_{i=1}^{N}\left(  \frac{\tilde{R}_{K}^{(i)}-y}{\delta\tilde
{R}_{K}^{(i)}}\right)  ^{2}
\text{ ,}
\label{chisq}
\end{equation}
with the experimental points and their errors given by $\tilde{R}_{K}^{(i)} \pm \delta\tilde
{R}_{K}^{(i)}$. 
The experimental results on kaon production in \textit{pp} scattering are shown in Fig. 1, where data are taken from Ref. \cite{Gazdzicki:1996pk} (three left points) and from the compilation of Ref. \cite{Stepaniak:2023pvo}, see also Refs. \cite{Gazdzicki:1991ih,NA61SHINE:2024eqz,NA61SHINE:2021wba,NA61SHINE:2017fne}.
We check the `null hypothesis' $H_0$ of no isospin breaking in \textit{pp} scattering ($\tilde{R}_{K}=1$ for $\alpha = \beta$) by setting $y=1$ in Eq. (\ref{chisq}), which leads to the numerical value $\chi^2(y=1) = 6.270$. The probability of getting a worse $\chi^2(1)$ for 9 d.o.f. amounts to $p(\chi^2(y=1) >  6.270) = 0.71 > 0.05$. Hence, isospin symmetry cannot be rejected for \textit{pp} reactions.

Performing a weighted average of data delivers $\tilde{R}_K = 1.030 \pm 0.038$, compatible with one, as expected. 
As it is well known, the central value of the weighted average corresponds to the fit to a constant function, thus to the minimum of Eq. (\ref{chisq}),
while the uncertainty is $\sqrt{2/\left(  d^2\chi^{2}/dy^2\right)  _{y_{0}}}$ with $y_0 = \tilde{R}_K = 1.030$.
The minimum corresponds to $\chi^2_{min} = 5.64$ with a worse-fit probability (for 8 d.o.f.) of $p(\chi^2_{min} > 5.64) = 0.69$, showing that a constant as a function of the scattering energy describes the data well; the $\chi^2_{min}$ per d.o.f. amounts to $0.94 \simeq 1$. 

Next, we turn to nucleus-nucleus results using the world data compilation in Ref. \cite{NA61SHINE:2023azp}, see Fig. 2.
Again, we first test the $H_0$ hypothesis: `isospin symmetry is not violated'. Setting $y=1$ in Eq. (\ref{chisq}) and using the experimental values of Fig. 2, one gets  $\chi^2(y=1) = 50.3$. This large value corresponds to a rather small worse-fit probability (for 15 d.o.f.): $p(\chi^2(y=1) > 50.3) = 1.1 \cdot 10^{-5}$. 

Treating the experimental values as uncorrelated, the weighted average leads to $\tilde{R}_K = 1.185 \pm 0.029$. Thus, we may conclude that $\tilde{R}_K$ is not compatible with unity at the $6.47 \sigma$-level: isospin symmetry is broken. Moreover, the violation is at the level of $18 \%$. 
Note, the corresponding $\chi^2_{min}= 8.39$ implies a $\chi^2$ per d.o.f. of $\chi^2(1.185)/14 \simeq 0.5 < 1$ \footnote{For uncorrelated results, the fact that the $\chi^2$ per d.o.f. is sizably below unity may point to overestimated uncertainties.}. 
The fit results for nucleus-nucleus scattering are summarized in Fig. 2. 
%
%

We summarize the findings above as such: \textit{pp} scattering data on kaon productions fulfill isospin, while nucleus-nucleus do not. It is then natural to speculate that the isospin breaking in the latter is caused by finite-density effects. In this respect, the very right point of Fig. 2 is interesting: this is the result of the ALICE collaboration \cite{ALICE:2013mez,ALICE:2013cdo}, which corresponds to a very small baryonic chemical potential. The central value of $\tilde{R}_K$ is close to one, but the error is too large to drive any conclusion.

If isospin is broken, the quantity $\tilde{R}_K$ depends on the nucleon-nucleon scattering energy as:
\begin{equation}
\tilde{R}_{K}=\frac{1\text{ }+2x\frac{1+Q/A}{2-Q/A}}{1\text{ }+\frac{2x}%
{r}\frac{1+Q/A}{2-Q/A}}\overset{\text{large }\sqrt{s_{NN}}}{\rightarrow}r = \frac{\alpha}{\beta}
\text{ ,} 
\end{equation}
where $r = \alpha/\beta$ is the $u/d$-ratio of quarks produced in the QCD vacuum, and $x=\alpha/n_{u}$ is the energy-dependent ratio of vacuum/initial $u$ quarks, that can be modeled as
$x=\lambda\left(  \sqrt{s_{NN}}\right)  ^{\kappa}$. Typically $\kappa \sim 0.3$-$0.5$ \cite{Gazdzicki:1998vd,Busza:2018rrf}. 
Note, $\tilde{R}_{K}=1$ for  $r=1$ for any energy.
The result of Fig. 2 shows that the present data break isospin but do not show any energy dependence. This means that an eventual energy dependence is hidden in the uncertainties and/or agrees with the situation in which the quarks generated by the QCD vacuum dominate. In any case, it is allowed to approximate $\tilde{R}_K \approx r = 1.185 \pm 0.029$.

\begin{figure}
    \centering
    \includegraphics[width=\linewidth,clip=true,trim=7pt 4pt 0 0]{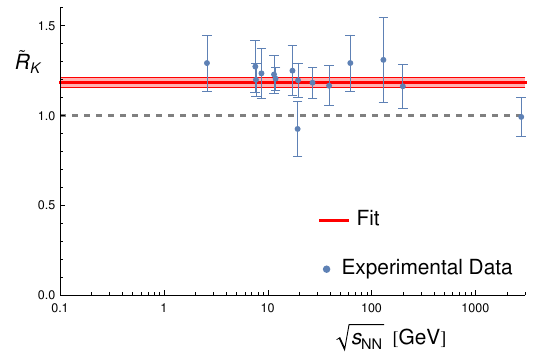}
    \caption{Experimental data for $\tilde{R}_K$ of Eq. (\ref{rkt}) from nucleus-nucleus collisions, together with the weighted average (constant fit) $\tilde{R}_K=1.185\pm0.029$. The dashed line is the isospin-symmetric prediction $\tilde{R}_K = 1$.}
    \label{fig:rktildeAA}
\end{figure}

\begin{figure}
    \centering
    \includegraphics[width=\linewidth,clip=true,trim=7pt 4pt 0 0]{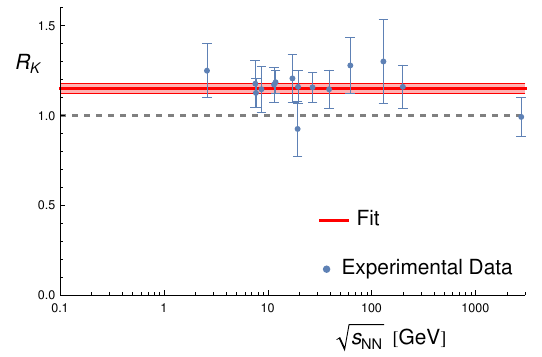}
    \caption{ Experimental data for $R_K$ of Eq. (\ref{rk1}) from nucleus-nucleus collisions, together with the weighted average (constant fit) $R_K = 1.152 \pm 0.027$. The dashed line is the isospin-symmetric prediction $R_K = 1$ valid for $Q/A = 1/2$.}
    \label{fig:rkAA}
\end{figure}

The previous result can be checked by looking at $R_{K}$, which in general, is also energy-dependent with
\begin{equation}
R_{K}=\frac{1\text{ }+2x}{\frac{2-Q/A}{1+Q/A}\text{ }+\frac{2x}{r}\text{ }%
}\overset{\text{large }\sqrt{s_{NN}}}{\rightarrow}r = \frac{\alpha}{\beta}
\text{ .}
\end{equation}
As Fig. 3 shows, the experimental data of Ref. \cite{NA61SHINE:2023azp} can be described by a constant $R_K= 1.152 \pm 0.027$, which is compatible with the result for $\tilde{R}_K$. The corresponding $\chi^2_{min} = 6.56$ (for $14$ d.o.f.) leads to $\chi^2_{min}/14 = 0.47 <1$.

Notably, in Ref. \cite{SeaQuest:2021zxb} it is shown that the proton contains more $\bar{d}$ valence quarks than $\bar{u}$ ones. At first, this result seems odd to the experimental findings of Figs. \ref{fig:rktildeAA} and \ref{fig:rkAA} summarized above. On the other hand, one needs to stress the differences among these systems: the distribution of valence quarks in the protons is not equivalent to the production of constituent quarks in heavy-ion collisions. In particular, the excess of $\bar{d}$ in the proton is not necessarily a manifestation of isospin breaking, because an important role may be played by the Pauli principle\footnote{One can argue that constraints due to the Pauli principle do not apply to newly created quark and anti-quark states in an open system as opposed to the nucleonic bound states.} \cite{Geesaman:2018ixo}. More specifically, charge-symmetry implies that an equal excess of $\bar{u}$ should occur for the neutron. Thus, an eventual breaking of charge-symmetry in parton distributions should be checked by averaging over charge-symmetric partners (such as the neutron and the proton).


Next, we turn to the consequence of isospin breaking. Namely, it should not affect only kaons, but also the multiplicities of other hadrons. We present estimates of some ratios in Table I and Table II for the case $Q/A \simeq 1/2$. The enhancement of the proton/neutron ratio of a factor of $r$ is interesting also for eventual cosmological implications, but it is not easy to measure because neutrons are not easy to detect. 
The hyperon ratio $\Sigma^{+}/\Sigma^{-} \approx  r^2$ is enhanced and involves long-lived charged particles, thus, it is of extreme interest. 
The $\Delta$-ratio $\Delta^{++}/\Delta^{-} = r^3$ is even more enhanced, but the $\Delta$-resonance lives short since it decays strongly. 

\begin{table}[h!]
\centering
\begin{tabular}
[c]{|l|l|}\hline
Ratio & Estimated value\\\hline
$R_{K}=\frac{K^{+}+K^{-}}{K^{0}+\bar{K}^{0}}$ & $r=1.185\pm0.029$\\\hline
$p/n$ & $r=1.185\pm0.029$\\\hline
$\pi^{+}/\pi^{0}$ & $\frac{2r}{1+r^{2}}=0.986\pm0.004$\\\hline
$\Sigma^{+}/\Sigma^{0}$ & $r=1.185\pm0.029$\\\hline
$\Sigma^{+}/\Sigma^{-}$ & $r^{2}=1.404\pm0.068$\\\hline
\end{tabular}
\caption{Multiplicity ratios for kaons, nucleons, pions, and hyperons.}
\label{table:1}
\end{table}

\begin{table}[h!]
\centering
\begin{tabular}
[c]{|l|l|}\hline
Ratio & Estimated value\\\hline
$\Delta^{++}/\Delta^{+}$ & $r=1.185\pm0.029$\\\hline
$\Delta^{++}/\Delta^{0}$ & $r^{2}=1.404\pm0.068$\\\hline
$\Delta^{++}/\Delta^{-}$ & $r^{3}=1.664\pm0.120$\\\hline
$\Sigma^{\ast+}/\Sigma^{\ast0}$ & $r=1.185\pm0.029$\\\hline
$\Sigma^{\ast+}/\Sigma^{\ast-}$ & $r^{2}=1.404\pm0.068$\\\hline
$\Xi^{\ast-}/\Xi^{\ast0}$ & $r=1.185\pm0.029$\\\hline
\end{tabular}

\caption{Multiplicity ratios of decuplet baryons.}
\label{table:1}
\end{table}

Finally, we extend the quark recombination approach to the case in which light antiquarks $\bar{u}$ and/or $\bar{d}$ are included in the initial
state (i.e., involving mesons and/or anti-nuclei). 
The probabilities are modified as follows:
\begin{align}
p(K^{+})  &  \propto n_{u}\gamma+\alpha\gamma\text{ ; }p(K^{-})\propto
n_{\bar{u}}\gamma+\alpha\gamma\text{  ,}\\
p(K^{0})  &  \propto n_{d}\gamma+\beta\gamma\text{; }p(\bar{K}^{0})\propto
n_{\bar{d}}\gamma+\beta\gamma
\text{ .}
\end{align}
The ratio \ $R_{K}$ emerges as
\begin{equation}
R_{K}=\frac{\left\langle K^{+}\right\rangle +\left\langle K^{-}\right\rangle
}{\left\langle 2K_{S}^{0}\right\rangle }=\frac{n_{u}+n_{\bar{u}}+2\alpha
}{n_{d}+n_{\bar{d}}+2\beta} \text{ .}
\label{eq:lightantiquarkRk}
\end{equation}
Again, $R_{K}=1$ in the isospin-symmetric limit ($\alpha=\beta$) is realized for
$n_{u}+n_{\bar{u}}=n_{d}+n_{\bar{d}}.$ 
This is the case for e.g. both $\pi^{-}C$
and $\pi^{+}C$ scattering (or for any nucleus with $Q/A=1/2$). Thus, isospin symmetry implies $R_K^{\pi^+ C} =R_K^{\pi^- C}  =1$. A departure from this value signalizes isospin-symmetry breaking in each of them separately. (Charge-symmetry implies only that $(R_K^{\pi^+ C} +R_K^{\pi^- C})/2  =1$.)  
Using the previous results for isospin breaking, we predict:
\begin{equation}
    R_K^{\pi^+ C} = R_K^{\pi^- C} \simeq 1.185 \pm 0.029
    \text{.}
\end{equation}
Very interestingly, in Ref. \cite{NA61SHINE:2022tiz}  $\pi^{-}C$ was studied, finding that $R_K\sim 1.2$, in line with the results of Figs. \ref{fig:rkAA} and  \ref{fig:rktildeAA}.

In the general case with initial (anti-)quarks $u$ and $d$ the appropriate $\tilde{R}_{K}$ takes the form:
\begin{align}
\tilde{R}_{K}  & =R_{K}+\frac{n_{d}+n_{\bar{d}}-n_{u}-n_{\bar{u}}}%
{n_{u}-n_{\bar{u}}}\frac{\left\langle K^{+}\right\rangle -\left\langle
K^{-}\right\rangle }{\left\langle 2K_{S}^{0}\right\rangle }\nonumber\\
& =\frac{n_{d}+n_{\bar{d}}+2\alpha}{n_{d}+n_{\bar{d}}+2\beta}
\label{rtkgen}
\end{align}
Just as before, $\tilde{R}_{K}=1$ in the isospin-symmetric limit ($\alpha = \beta$). This expression can be used to test isospin symmetry in scattering involving non-strange mesons, antiprotons, and also anti-nuclei.
Notice that Eq. 
(\ref{rtkgen}) 
is valid also for initial states with 
$n_s = n_{\bar{s}}$, thus for hidden-strange mesons, such as the $\eta$, $\eta^{\prime}$, and $\phi$, and for scattering with overall zero strangeness, such as $K^+ \Lambda$.

An additional interesting application of the quark coalescence model amounts to $R_K$ in $e^+$-$e^-$ scattering. Interpreting it as the ensemble of $n_u =n_{\bar{u}}=1$ with probability $4/6$, as well as $n_d =n_{\bar{d}}=1$ and $n_s =n_{\bar{s}}=1$ with probability $1/6$ each (we neglect heavier quarks), the ratio $R_K$ reads:
\begin{equation}
    R_K^{e^+ e^-} = \frac{4}{6} \frac{2+2\alpha}{2\beta} + \frac{1}{6} \frac{2\alpha}{2+2\beta} +\frac{1}{6} \frac{\alpha}{\beta}  
    \text{ ,}
\end{equation}
as it follows from Eq.~(\ref{eq:lightantiquarkRk}).
Since the absolute values $\alpha$ and $\beta$ are unknown, $R_K$ cannot be univocally predicted. On top, an energy dependence is necessarily present for this ratio.  Yet, from the above equation, it follows that $R_K \geq \alpha/\beta = r$, where $r$ is the asymptotic value at high energy when the quarks produced by the QCD vacuum dominate. The question is if this case is similar to $pp$ (no isospin breaking, $r=1$) or to nucleus-nucleus one $r \simeq 1.2$. A detailed study is left as an outlook.  
For recent experimental results, see Ref. \cite{BESIII:2025mbc}.

In the most general case with arbitrary $n_{u,d,s}$ and $n_{\bar{u},\bar{d},\bar{s}}$ the quantity $\tilde{R}_K$ reads 
\begin{equation}
\tilde{R}_{K}=\frac{\left(  n_{d}+\alpha\right)  \left(  n_{\bar{s}}%
+\gamma\right)  +\left(  n_{\bar{d}}+\alpha\right)  \left(  n_{s}%
+\gamma\right)  }{\left(  n_{d}+\beta\right)  \left(  n_{\bar{s}}%
+\gamma\right)  +\left(  n_{\bar{d}}+\beta\right)  \left(  n_{s}%
+\gamma\right)  }\nonumber
\text{ .}
\end{equation}
However, it cannot be expressed as a function of the three multiplicities $\langle K^+ \rangle$, $\langle K^- \rangle$, and $\langle K_S^0 \rangle$, but it involves separately $\langle K_0 \rangle$
 and $\langle \bar{K}_0 \rangle$ \footnote{The expression is lengthy. Since it cannot be used in practice, we do not report it here.}. 
 This fact is not convenient because only $K_S^0$ is usually detected. Moreover, even measuring $K_L^0$ would not help, since (neglecting a very small $CP$-breaking) $\langle K_L^0 \rangle= \langle K_S^0 \rangle$, implying that the multiplicities  $\langle K_0 \rangle$ and
 $\langle \bar{K}_0 \rangle$ cannot be obtained.

 In conclusion, within a quark recombination approach, we have introduced a modified ratio of kaon multiplicities $\tilde{R}_K$ (eq. (\ref{rkt})), which is unity when isospin is conserved, independently of the scattering energy and the employed nuclei or nucleons.  We confirm that the kaon multiplicities in heavy-ion collisions display a \textit{large} breaking of isospin symmetry (at the $6.4\sigma$-level): substantially more $u$ than $d$ quarks are produced. This is at odds with proton-proton scattering, where isospin is conserved, thus suggesting a finite density effect inherent to the collision of nuclei. 
 Also, the possibility that $Q/A$ is not always a strict constant because of the eventual inhomogeneous distribution of protons and neutrons within nuclei may lead to $(Q/A)_{eff}$ that is worth investigating as an outlook. 
In the future, more precise data may also allow for the determination of the energy dependence of both $R_K$ and $\tilde{R}_K$. 
An interesting case is the choice of nuclei with $Q=A/2$, for which $R_K =  \tilde{R}_K$. In this respect, oxygen-oxygen and carbon-carbon seem feasible. Also, deuteron-deuteron scattering can be valuable as the smallest system with nuclei with $Q=A/2$. Is in this case $R_K=\tilde{R}_K$ compatible with unity?
The information above can be combined with other sources of information, such as the electron-positron
scattering \cite{BESIII:2025mbc}.

The generalization to antiquarks in the initial state has been put forward ($\tilde{R}_K$ in Eq. (\ref{rtkgen})). It can help to interpret pion-nucleus data, and it may be used for certain reactions involving $s$-quarks.  
Finally, predictions for other enhanced multiplicity ratios, such as isospin-violating proton/neutron and hyperon $\Sigma^+ / \Sigma^-$  ones, are promising observables for future experiments. 

\bigskip

\textbf{Acknowledgments}: The authors thank M. Gaździcki, A. Rybicki, S. Mrówczyński, M. Rybczyński, P. Man Lo, L. Tinti, S. Samanta, and
K. Grebieszków for very useful discussions.

\bibliography{Ref}
\bibliographystyle{unsrt}

\end{document}